\documentclass[1p,dvips]{elsarticle}
\usepackage{latexsym}
\usepackage{amsmath}
\usepackage{graphicx}
\usepackage{subfigure}

\begin{document}
\begin{frontmatter}
\title{Environmental effects on the $x^{4}$ model in Tsallis statistics}
\journal{}
\author{Masamichi Ishihara\corref{cor}}
\cortext[cor]{Corresponding author. Tel.: +81 24 932 4848; Fax: +81 24 933 6748.}
\ead{m\_isihar@koriyama-kgc.ac.jp}
\address{Department of Human Life Studies, Koriyama Women's University, Koriyama, 963-8503, Japan}

\begin{abstract}
The author studied the effects of the environment described by Tsallis statistics in quantum mechanics, 
when the deviation from Boltzmann-Gibbs (BG) statistics is small. 
The $x^{4}$ model was used and 
the squeeze angle caused by the difference between Tsallis and BG statistics 
was calculated perturbatively  in the mean field approximation 
as a function of the dimensionless parameters: 
the inverse temperature $\beta_p$ and the coupling strength $\lambda_p$.
The author found that the effect of the deviation from BG statistics is relatively large 
at high temperature. 
The squeeze angle as a function of $\beta_p$ has a dip structure, 
and the dip is deeper with the increase of $\lambda_p$.
The angle as a function of $\beta_p$ changes the sign.  
These facts indicate that 
the frequency is modulated by the difference between these statistics.
\end{abstract}
\begin{keyword}
Non-extensive statistics; Squeezing; $x^4$ model
\end{keyword}
\end{frontmatter}

\section{Introduction}
An extended equilibrium statistics, so-called Tsallis statistics \cite{Tsallis}, 
was introduced and have been investigated in the past few decades.
This statistics shows a power-law distribution and a non-additive entropy.
A weighted integral of Gauss distribution (superstatistics) is equal to a Tsallis distribution
\cite{Queiros08}.
It was shown that a certain Langevin equation leads to a Tsallis distribution
\cite{Fuentes}.
These facts imply that the non-extensivity should appear generally. 
Some physical origins of Tsallis distribution were suggested \cite{Wilk07_Tsallis}.
For an example, long-range forces may cause power-law distributions.
The distribution of 
such a system whose ingredients interact with each other through long-range forces, 
may be different from that in the Boltzmann-Gibbs (BG) statistics. 
Then some environments should be described by Tsallis statistics.

This non-extensive statics has been widely applied to various phenomena and methods, 
such as 
particle distribution at high energies \cite{Wilk07_Tsallis},
non-extensive network \cite{Hasegawa_physica}, 
generalized simulated annealing algorithm\cite{Andricioaei_PRE53} and so on. 
This statistics has two parameters: one is the inverse of the temperature $\beta$ and 
the other is the parameter $q$ which indicates 
the difference between Tsallis statistics and BG statistics.
The parameter $q$ obtained by fitting the particle distribution \cite{Wilk07_Tsallis}
is close to the value $q=1$ 
at which the distribution corresponds to BG distribution.

The effects of the environment described by Tsallis statistics in Quantum Field Theory
were also studied.  
An example is the calculation of gluon propagator in the environment\cite{Kohyama_Tsallis06}. 
It is well-known that 
the propagator in the thermal environment described by BG statistics is modified, 
and then the propagator in the environment described by Tsallis statistics is also modified.
Similarly, 
it is expected that the expectation values of other physical quantities are affected by 
the environment.

The study of the modification of a physical quantity
by the environment described by Tsallis statistics 
is important to clarify the effects of the Tsallis distribution, 
because power-like distributions appear in many systems. 
The absolute value $|1-q|$ (the index of the deviation from BG statistics) is small 
in many cases, as shown in other studies. 
Therefore I attempt to calculate the physical quantities, 
such as the expectation value of the square of the coordinate, 
under the influence of the environment described by Tsallis statistics,  
when the deviation from the BG statistics is small enough. 
I use the $x^{4}$ model to study the effects of the environment 
through the anharmonic potential. 
I apply a mean field approximation and Bogoliubov transformation. 
The energy depends on the dimensionless parameters, 
the inverse temperature $\beta_p$ and the coupling strength $\lambda_p$, 
and is calculated in the self-consistent manner.
The effects of the deviation from BG statistics are given numerically 
by estimating the squeeze angle.
I found that the squeeze angle as a function of $\beta_p$ has a dip structure 
and the angle changes the sign.  
The frequency is modulated by the difference between these statistics,
because the coefficient of the $x^2$ term is modified.
These results provide the insight in the study of the phenomena described by the model 
with self-coupling, such as the $\phi^4$ theory.

\section{Squeeze angle}
\label{sec:squeeze_Tsallis}
\subsection{Squeeze angle in a mean field approximation}
\label{sec:squeeze_parameter}
I deal with the $x^{4}$ model in the present paper. 
The Hamiltonian is 
\begin{equation}
H = \frac{1}{2m} \left( p^{2} + m^{2} \omega^{2} x^{2} \right) + \frac{1}{4!} m \omega^{2} \lambda x^{4} ,
\end{equation}
where $x$ is the coordinate, $p$ is the momentum, $m$ is the mass,
$\omega$ is the frequency and $\lambda$ is the coupling strength. 
The mean field approximation \cite{Boyanovsky97_Scalar,MeanField} is applied:
\begin{subequations}
\begin{align}
& x^{2n} \rightarrow 
\frac{(2n)!}{2^n (n-1)!} \ \left( \langle x^{2} \rangle_{q} \right)^{n-1} \ x^{2} - 
\frac{(n-1) [(2n)!]}{2^n n!} \ \left( \langle x^{2} \rangle_{q}\right)^{n} , \\ 
& x^{2n+1} \rightarrow 
\frac{(2n+1)!}{2^n n!} \ \left( \langle x^{2} \rangle_{q} \right)^{n} \ x , 
\end{align}
\end{subequations}
where 
$\left\langle O \right\rangle_{q}$ is the statistical average of a physical quantity $O$
in Tsallis statistics.
The mean field Hamiltonian is given by 
\begin{align}
H^{\mathrm{MF}} 
 = \frac{1}{2m} \left( p^{2} + M^{2}(q,\beta) \omega^{2} x^{2} \right)
 - \frac{1}{8} m \omega^{2} \lambda \left( \left\langle x^{2} \right\rangle_{q} \right)^{2}
,
\label{eqn:MeanFieldHamiltonian}
\end{align}
where $\beta$ is the inverse of the temperature and  
$M^{2}(q,\beta)$ is defined as $m^{2} \left[1 + \frac{\lambda}{2} \left\langle x^{2} \right\rangle_{q} \right]$. 
The coefficient of the $x^2$ term is influenced by the environment \cite{Gavin} 
through $\left\langle x^{2} \right\rangle_{q}$. 
I introduce an arbitrary mass $M$ and define creation and annihilation operators, 
$a^{\dag}$ and $a$, which are related to $x$ and $p$: 
\begin{align}
x = \sqrt{ \frac{\hbar}{2M\omega} } \left( a^{\dag} + a\right), 
\qquad 
p = i \sqrt{ \frac{M\omega \hbar}{2} } \left( a^{\dag} - a \right).
\label{eqn:annihilation}
\end{align}
With eqs.\eqref{eqn:annihilation}, 
the mean field Hamiltonian is given by 
\begin{align}
H^{\mathrm{MF}} 
&= \frac{M}{m} \frac{\hbar \omega}{2} \left\{
\left(1+\frac{M^{2}(q,\beta) - M^{2}}{2M^{2}}\right)\left(a a^{\dag} + a^{\dag} a \right) 
\right. \nonumber \\ & \quad \left. 
+ \left(1+\frac{M^{2}(q,\beta) - M^{2}}{2M^{2}}\right)\left(a^{2} + \left(a^{\dag}\right)^{2} \right)  
\right\}
- \frac{1}{8} m\omega^{2} \lambda \left( \left\langle x^{2} \right\rangle_{q} \right)^{2}
.
\end{align}

The Bogoliubov transformation \cite{Walls} is applied to diagonalize $H^{\mathrm{MF}}$:
\begin{align}
a = \cosh \theta \cdot a_{q} + \sinh \theta \cdot a_{q}^{\dag},
\qquad 
a^{\dag} = \cosh \theta \cdot a_{q}^{\dag} + \sinh \theta \cdot a_{q}, 
\label{eqn:relation:a:a_q}
\end{align}
where the parameter $\theta$ is the squeeze angle.
The quantum state $|0,q;\beta \rangle$ is defined by 
the equation, $a_{q} |0,q;\beta\rangle = 0$.
Here, the function $g(q,\beta)$ is defined by 
\begin{equation}
g(q,\beta) := \frac{M^{2}(q,\beta) - M^{2}}{2M^{2}} . 
\label{def:g_of_q}
\end{equation}
The mean field Hamiltonian is diagonalized, 
if the following equation is satisfied: 
\begin{equation}
\tanh (2\theta(q,\beta)) =  - \frac{g(q,\beta)}{1+g(q,\beta)}. 
\label{eqn:1:theta}
\end{equation}
The mean field Hamiltonian is given by
\begin{equation}
H^{\mathrm{MF}} =
\frac{\Omega(q,\beta)}{2}
\left(a_{q} a_{q}^{\dag}  + a_{q}^{\dag}  a_{q} \right)
- \frac{1}{8} m\omega^{2} \lambda \left( \left\langle x^{2} \right\rangle_{q} \right)^{2}
,
\end{equation}
where $\Omega(q,\beta)$ is defined by  
\begin{equation}
\Omega(q,\beta) := \left(\frac{M}{m}\right) \hbar \omega \left(1+\frac{g(q,\beta)}{1+g(q,\beta)}\right) \cosh(2\theta) .
\label{eqn:def:omegaq}
\end{equation}

I show another way to obtain eq.~\eqref{eqn:1:theta}.
Similarly to eq.~\eqref{eqn:annihilation}, 
the operators, $x$ and $p$, are related to $a_q$ and $a_q^{\dag}$:
\begin{align}
x = \sqrt{ \frac{\hbar}{2M(q,\beta)\omega} } \left( a_q^{\dag} + a_q \right) , 
\qquad 
p = i \ \sqrt{ \frac{M(q,\beta) \omega \hbar}{2} } \left( a_q^{\dag} - a_q \right) .
\label{eqn:annihilation:q}
\end{align}
From eqs.~\eqref{eqn:annihilation},~\eqref{eqn:relation:a:a_q}, and \eqref{eqn:annihilation:q},
I obtain the following equation:
\begin{equation}
\exp\left(\theta\right) = \sqrt{ \frac{M}{M(q,\beta)} }.
\label{eqn:2:theta}
\end{equation}
Apparently, this equation satisfies eq.~\eqref{eqn:1:theta}.
Then I define $\theta_{ij}$ as follows:
\begin{equation}
\exp \left(\theta_{ij}\right) := \sqrt{ \frac{M_i}{M_j} } .
\label{def:general:theta}
\end{equation}
Equation~\eqref{def:general:theta} indicates that 
\begin{equation}
\theta_{13} = \theta_{12} + \theta_{23} .
\label{eqn:combination}
\end{equation}
This relation is useful to find the value of the parameter $\theta$. 

The Hamiltonian is reduced to be harmonic in the mean field approximation.
The expressions of $\left\langle x^{2} \right\rangle_{q}$ and $\left\langle p^{2} \right\rangle_{q}$ for the Hamiltonian are given by
\begin{align}
\left\langle x^{2} \right\rangle_{q}
= \left( \frac{\hbar}{2M\omega} \right) e^{2\theta} 
   \left( 2 \left\langle a_{q}^{\dag} a_{q} \right\rangle_{q} + 1 \right) 
 , \qquad 
\left\langle p^{2} \right\rangle_{q} 
= \left( \frac{M \hbar \omega}{2} \right) e^{-2\theta} 
   \left( 2 \left\langle a_{q}^{\dag} a_{q} \right\rangle_{q} + 1 \right)
,
\end{align}
where the equation $\left\langle a_{q} \right\rangle_{q} = 0$ is satisfied.
Here I note that the mass $M$ is not specified in the previous discussion. 

\subsection{The squeeze angle caused by the difference between Tsallis and Boltzmann-Gibbs statistics}
In Tsallis statistics, 
the statistical average of a physical quantity $O$ is given by
\begin{equation}
\langle O \rangle_{q} = 
\frac{\displaystyle\sum_{i} 
\big\langle i \big| O
\left\{ \left[ 1 - (1-q) \frac{\beta}{c_q} (H - \langle H \rangle_{q} )\right]^{\frac{q}{1-q}}  \right\} \big| i \big\rangle }
{\displaystyle\sum_{j} \big\langle j \big|  \left[ 1 - (1-q) \frac{\beta}{c_q} (H - \langle H \rangle_{q})\right]^{\frac{q}{1-q}} \big| j \big\rangle}
\label{eqn:expectation_in_Tsallis}
,
\end{equation}
where 
$\big| i \big\rangle$  is a quantum state labelled $i$, 
$c_q$ is the normalization factor
and $\langle H \rangle_{q}$ is the expectation value of the Hamiltonian.

I use the approximated Hamiltonian $H^{\mathrm{MF}}$ instead of  $H$ 
in eq.~\eqref{eqn:expectation_in_Tsallis}.
To simplify, I introduce the Hamiltonian $\tilde{H}$ shifted as 
$\tilde{H}= H^{\mathrm{MF}} - \langle 0, q; \beta | H^{\mathrm{MF}} | 0, q; \beta \rangle$. 
When the quantity $O$ is the Hamiltonian $\tilde{H}$, 
I obtain the self-consistent equation for $\tilde{H}$
by using the number state $|n,q; \beta \rangle$ 
which is the $n$ particle state constructed with $a_{q}^{\dag}$ on the vacuum $|0,q; \beta \rangle$:
\begin{equation}
\langle \tilde{H} \rangle_{q} = 
\frac{\displaystyle\sum_{n=0}^{\infty} \left\{ \Omega(q,\beta)\ n \left[ 1 - (1-q) \frac{\beta}{c_q} (\Omega(q,\beta)\ n - \langle \tilde{H} \rangle_{q} )\right]^{\frac{q}{1-q}} \right\}}
{\displaystyle\sum_{n=0}^{\infty} \left[ 1 - (1-q) \frac{\beta}{c_q} (\Omega(q,\beta)\ n - \langle \tilde{H} \rangle_{q})\right]^{\frac{q}{1-q}}}
. 
\label{eqn:tildeH}
\end{equation}
In this paper, I attempt to find the quantum state (namely squeeze angle $\theta$) 
in Tsallis statistics with small deviation from $q=1$.
Conventionally, the deviation $\epsilon$ from $q=1$ is defined by $\epsilon := 1-q$.
I use this parameter $\epsilon$ instead of $q$ in the following calculations.
I assume that 
the lowest contribution of $\theta$ for small $\epsilon$ is proportional to $\epsilon$. 
The functions, $\Omega(q,\beta)$, $c_{q}$ and $\theta(q,\beta)$, are expanded as series of $\epsilon$:
\begin{subequations}
\begin{align}
& \Omega(q,\beta) \equiv \Omega(1-\epsilon,\beta) = \Omega_0 - \epsilon \Omega_1 + O\left(\epsilon^{2}\right) \label{eqn:expansion_Omega},
\\
& c_{q} \equiv c_{1-\epsilon} = c_0 - \epsilon c_1 + O\left(\epsilon^{2}\right) ,
\label{eqn:expansion_Normalization}
\\
& \theta(q,\beta) \equiv \theta(1-\epsilon,\beta) = \theta_{0} - \epsilon \ \Theta + O\left(\epsilon^{2}\right) .
\label{eqn:expansion_of_theta}
\end{align}
\end{subequations}

I specify the mass $M$ explicitly to calculate the parameter $\theta$. 
I choose $M(q=1,\beta)$ as $M$.  
To avoid the confusion, I define the symbol $M_{\beta}$ by $M_\beta = M(q=1,\beta)$. 
In such the case, the angle $\theta$ satisfying eq.~\eqref{eqn:1:theta} reaches zero 
as $\epsilon \rightarrow 0$.
Therefore, the term $\theta_0$ in eq.~\eqref{eqn:expansion_of_theta} is equal to zero
in the case that $M$ is equal to $M_\beta$.

The expectation value $\langle \tilde{H} \rangle_{q}$ is also a function of $\epsilon$.
I expand this expectation as follows:
\begin{equation}
\langle \tilde{H} \rangle_{q} \equiv \langle \tilde{H} \rangle_{1-\epsilon} 
= E_{0} - \epsilon E_{1} + O(\epsilon^{2})
\label{eqn:expansion_tilde_H}
\end{equation}
The expressions of $c_0$ and $c_1$ can be obtained by using the relation between 
$c_q$ and the partition function $Z_q$:
\begin{equation}
c_q = \left( Z_q \right)^{1-q}. 
\label{eqn:nomarlization_Zq}
\end{equation}
In the similar way, $Z_q$ is expanded as 
$Z_{1-\epsilon} = Z_0 - \epsilon Z_1 + O(\epsilon^{2})$,
where $Z_0$ is the partition function at $q=1$. 
From eq.~\eqref{eqn:nomarlization_Zq}, these quantities are given by
\begin{align}
c_0 = 1, \qquad 
c_1 = -\ln Z_0.
\end{align}
The quantities, $E_0$ and $E_1$, are obtained by substituting 
eqs.~\eqref{eqn:expansion_Omega}, \eqref{eqn:expansion_Normalization}, \eqref{eqn:expansion_tilde_H} into eq.~\eqref{eqn:tildeH}.
The quantity $E_0$ is given by 
\begin{equation}
E_0 = \Omega_0 \left( \frac{S_1}{S_0} \right) ,
\end{equation}
where $S_p$ is defined by 
\begin{equation}
S_p := 
\sum_{n=0}^{\infty} n^{p} \exp \left[ - \beta \left( \frac{\Omega_0}{c_0} \right) n \right] .
\end{equation}
The quantity $\Omega_0$ is apparently given by 
\begin{equation}
\Omega_0 = \left( \frac{M_\beta}{m} \right) \ \hbar \omega .
\end{equation}
I obtain easily the expression of $\left\langle x^{2} \right\rangle_{q=1-\epsilon}$
by the perturbation with respect to $\epsilon$:\begin{align}
\left\langle x^{2} \right\rangle_{q=1-\epsilon}
= &
\frac{\hbar}{2 \omega M_{\beta}} R_{1} 
\nonumber \\ &
+\frac{\epsilon \hbar}{\omega M_{\beta}}
\left\{
- \Theta R_{1} 
+ \left[
\beta \left( 1 - \frac{c_1}{c_0} \right)\left( \frac{\Omega_0}{c_0} \right)
+ \beta \left( \frac{\Omega_1}{c_0} \right)
+ \beta^{2} \left( \frac{\Omega_0}{c_0} \right) \left( \frac{E_0}{c_0} \right)
\right] R_{2} 
\right. \nonumber \\ & \left.
+ \frac{\beta^{2}}{2} \left( \frac{\Omega_0}{c_0} \right)^{2} R_{3}
\right\}
+ O\left(\epsilon^{2}\right).
\label{eqn:x2:epsilon}
\end{align}
The functions $R_1$, $R_2$ and $R_3$ are defined by 
\begin{align}
R_1 := 1 + 2 \left( \frac{S_1}{S_0} \right) , \quad  
R_2 := \left( \frac{S_1}{S_0} \right) \left[ \left( \frac{S_2}{S_1} \right) - \left( \frac{S_1}{S_0} \right) \right] , \quad  
R_3 := \left( \frac{S_1}{S_0} \right) \left[ \left( \frac{S_2}{S_0} \right) - \left( \frac{S_3}{S_1} \right) \right] .
\end{align}
Apparently, 
the quantity $\langle x^{2} \rangle_q$ - $\langle x^{2} \rangle_{q=1}$ is $O(\epsilon)$.
Then the order of $g(1-\epsilon,\beta)$ is $O(\epsilon)$.
By using the definition of $g(q,\beta)$ and eq.~\eqref{eqn:x2:epsilon}, 
the expression of $g(1-\epsilon,\beta)$ for small $\epsilon$ is given as follows:
\begin{align}
& g(1-\epsilon,\beta) 
\nonumber \\ &
= \epsilon \ \frac{
\kappa \left\{ 
- R_1 \Theta +
\left[ \beta \left(1 - \frac{c_1}{c_0} \right) \left( \frac{\Omega_0}{c_0}\right) 
  + \beta \left( \frac{\Omega_1}{c_0} \right)
  + \beta^{2} \left( \frac{\Omega_0}{c_0}\right) \left( \frac{E_0}{c_0}\right)  \right] R_2 
  + \frac{1}{2} \beta^{2} \left( \frac{\Omega_0}{c_0} \right)^{2} R_3 
\right\}
}{1 + \kappa R_1}
\nonumber \\ &\quad
+ O(\epsilon^{2}) , 
\label{eqn:g_1-epsilon}
\end{align}
where $\kappa = ({\lambda \hbar})/({4 \omega M_{\beta} })$.
Therefore, eq.~\eqref{eqn:def:omegaq} gives the equation 
between $\Omega_1$ and $g(1-\epsilon,\beta)$ to the order $\epsilon$
by ignoring $O(\epsilon^2)$ terms:
\begin{equation}
\epsilon \Omega_{1} = - \Omega_0 g(1-\epsilon,\beta).
\label{eqn:rel:Omega1:g}
\end{equation}
Similarly, 
eq.~\eqref{eqn:1:theta} gives us the equation to the order $\epsilon$:
\begin{equation}
2 \epsilon \Theta = g(1-\epsilon,\beta).
\label{eqn:Theta:g}
\end{equation}
I finally obtain the expression of $\Theta$ with eqs.~\eqref{eqn:g_1-epsilon} 
\eqref{eqn:rel:Omega1:g} and \eqref{eqn:Theta:g}:
\begin{equation}
\Theta = \frac{
\kappa \left\{ 
\left[ \beta \left( 1- \frac{c_1}{c_0} \right) \left( \frac{\Omega_{0}}{c_0} \right)
     + \beta^{2} \left(\frac{\Omega_0}{c_0}\right) \left( \frac{E_0}{c_0}\right) \right] R_{2} 
+ \frac{1}{2} \beta^{2} \left( \frac{\Omega_0}{c_0} \right)^{2} R_3 \right\}
}{
2 + 3 \kappa R_1 + 2 \kappa \beta \left( \frac{\Omega_0}{c_0}\right) R_2 
} ,
\label{eqn:Theta}
\end{equation}
where $c_1$ is given by 
\begin{equation}
c_1 = -\ln Z_0 
    = -\ln S_0 - \beta \left( \frac{\Omega_0}{c_0} \right) \left( \frac{S_1}{S_0} \right) . 
\end{equation}
The function $S_p$ is simplified with the functions $f(x)$ and $\tilde{f}(x)$ which are
defined by   
\begin{align}
f(x) := \frac{e^{x}}{e^{x}-1}, \qquad 
\tilde{f}(x) := \frac{1}{e^{x}-1}.
\end{align}
For an example, 
the quantity ${S_1}/{S_0}$ is equal to $\tilde{f}\left( \beta \Omega_0 / c_0 \right)$.

The squeezing also occurs due to the difference between $M_{\beta}$ and $m$. 
The corresponding squeeze angle is given by 
\begin{equation}
\exp\left(\varphi_m(\beta) \right) = \left( \frac{m}{M_{\beta}} \right)^{1/2} .
\label{def:squeeze_param:temperature_only}
\end{equation}
The quantity $M(q=1,\infty)$ is not equal to $m$ when $\lambda$ is not zero.
Therefore $\varphi_m(\infty)$ is not zero generally.
From eq.~\eqref{eqn:combination}, 
the total squeeze angle $\varphi_{\mathrm{tot}}$ of the order $\epsilon$ is given by 
\begin{equation}
\varphi_{\mathrm{tot}} = \varphi_m(\beta) + \theta(1-\epsilon,\beta)
                      = \varphi_m(\beta) - \epsilon \Theta +O\left(\epsilon^{2}\right).
\end{equation}
The angle $\varphi_{\mathrm{tot}}$ is calculated 
by solving the temperature dependence of $M_{\beta}$ and using eq.~\eqref{eqn:Theta}.

\section{Numerical Estimation of the Squeeze Angle}
\label{sec:Numerical_Estimation}
In this section, I calculate numerically the angle $\Theta$ given by eq.~\eqref{eqn:Theta}
and the angle $\varphi_m(\beta)$ given by eq.~\eqref{def:squeeze_param:temperature_only}.
I introduce the dimensionless parameters, $\beta_p$ and $\lambda_p$, by 
\begin{align}
\beta_p = \beta \hbar \omega, 
\qquad 
\lambda_p = \lambda \frac{\hbar}{m\omega}.
\end{align}
The parameters in these numerical calculations, $\hbar$, $\omega$ and $m$, are set to 1.

\begin{figure}
\subfigure[The value of $\varphi_m$]{
\includegraphics[width=0.45\textwidth]{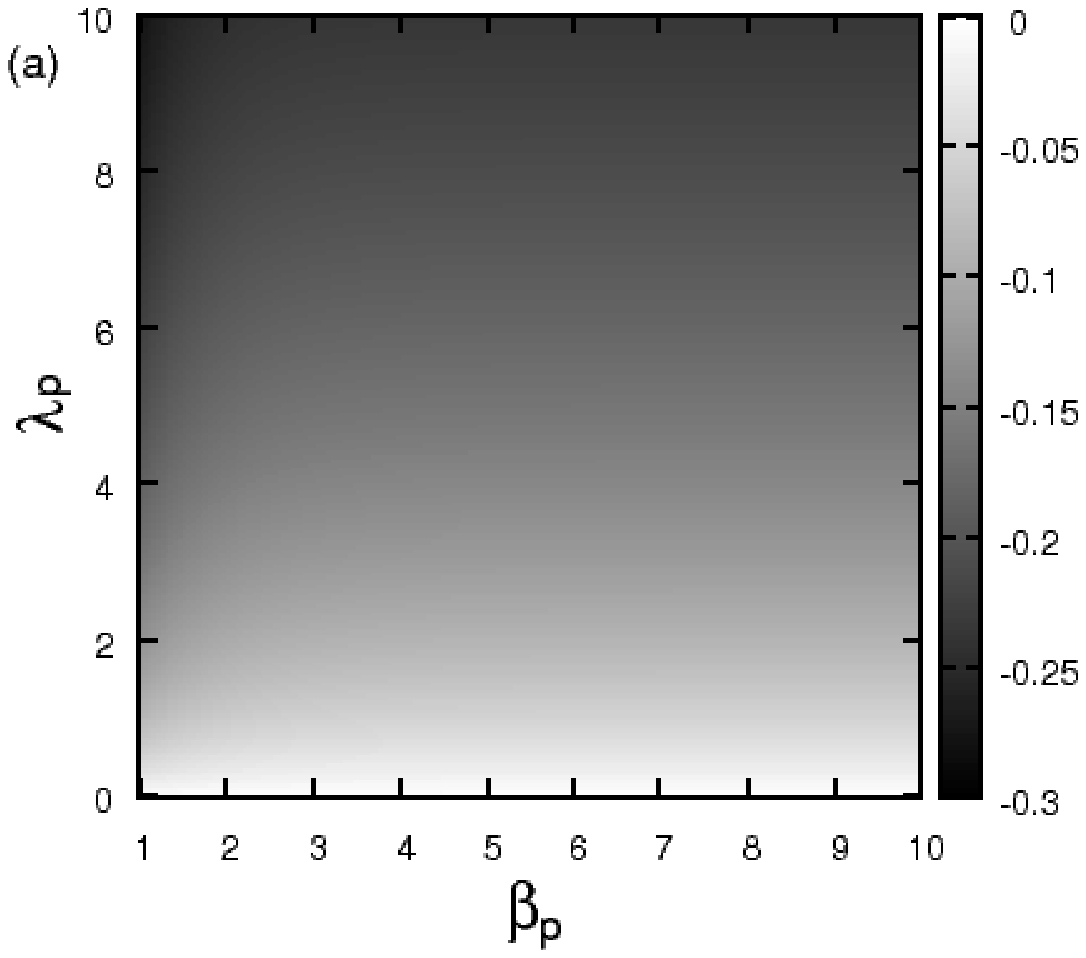}
\label{Fig:Phim_B1.0-10_L0-10_Thermal}
}
\subfigure[The value of $\Theta$]{
\includegraphics[width=0.45\textwidth]{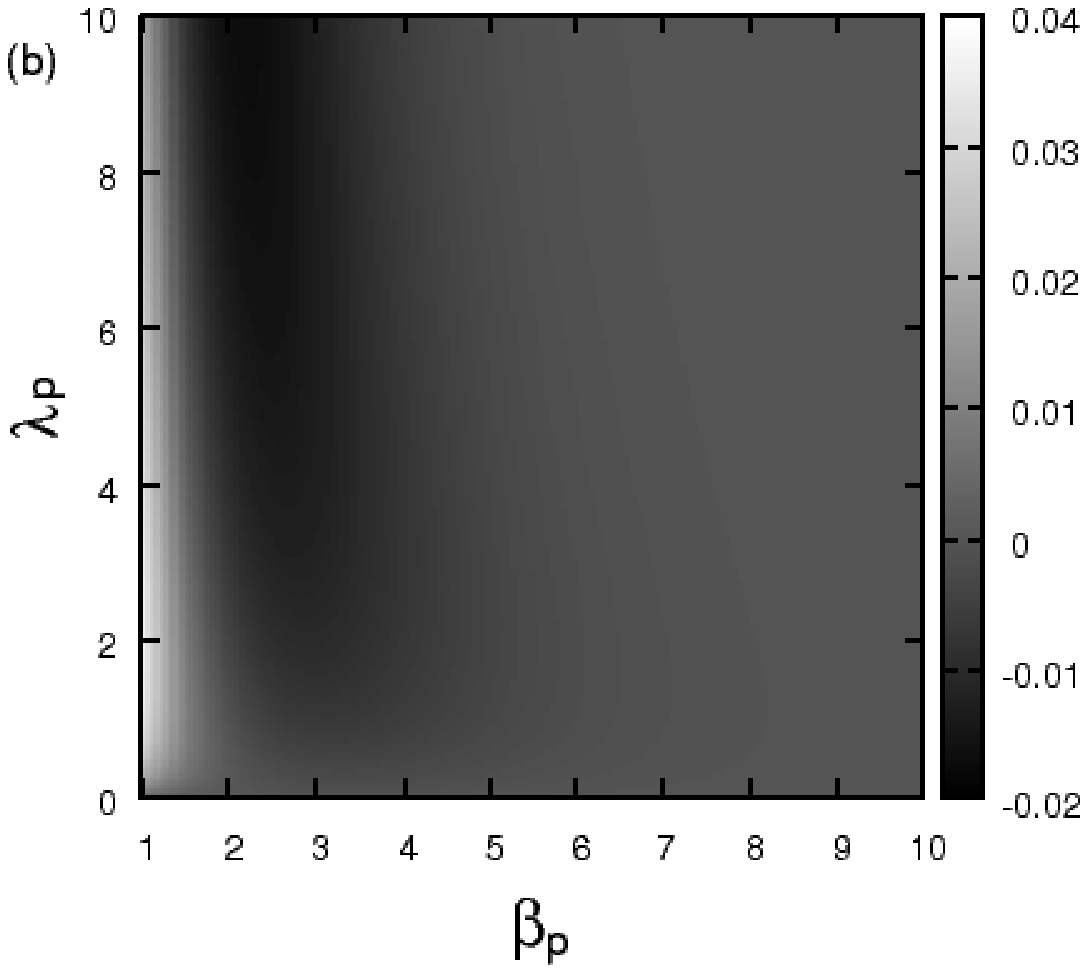}
\label{Fig:Theta_B1.0-10_L0-10_bk}
}
\caption{The squeeze angle in the ranges of $1.0 \le \beta_p \le 10$ and $0 \le \lambda_p \le 10$.}
\label{Fig:Theta_Phim:B1.0-10_L0-10}
\end{figure}

I show the values of $\varphi_m$ and $\Theta$ at various values of $\beta_p$ and $\lambda_p$. 
Figure~\ref{Fig:Phim_B1.0-10_L0-10_Thermal}  is the map of $\varphi_m$ 
in the ranges of $ 1 \le \beta_p \le 10$ and $0 \le \lambda_p \le 10$.
As is well-known, 
the temperature-dependent mass $M_{\beta}$ increases generally with the temperature,
because of the existence of the term  $m^{2} \lambda \langle x^{2} \rangle_{q=1} /2$.
Then $\varphi_m$ is negative at finite temperature and decreases with the temperature 
$\beta_p^{-1}$ and the coupling $\lambda_p$, 
as shown in Fig.~\ref{Fig:Phim_B1.0-10_L0-10_Thermal}.
Figure~\ref{Fig:Theta_B1.0-10_L0-10_bk} shows the map of $\Theta$ in the same range.
The value of $\Theta$ goes to zero from negative value as $\beta_p$ goes to infinity, 
as seen in this figure. 
This behavior is reasonable 
because the parameter $\beta_p$ corresponds to the inverse of the temperature.
As shown in this figure, 
$\Theta$ has a local minimum as a function of $\beta_p$ at a fixed $\lambda_p$ and 
the sign of $\Theta$ changes in the vicinity of $\beta_p = 2 \sim 3$,
while $\varphi_m$ decreases monotonically and the sign of $\varphi_m$ is negative.

The sign of $\Theta$ depends on the values of $\beta_p$ and $\lambda_p$, 
and this fact indicates that the sign of 
$\theta = - \epsilon \Theta$ does not depend on only the sign of 
the parameter $\epsilon$. 
In other words, the sign of $\theta$ cannot be determined by only the property of statistics, 
subextensive ($\epsilon <0$) or superextensive ($\epsilon>0$).
From eq.~\eqref{eqn:2:theta} and \eqref{eqn:expansion_of_theta}, I have
\begin{equation}
M(q=1-\epsilon,\beta) = M_{\beta} \exp\left(2 \epsilon \Theta \right).
\end{equation}
Then $M(q,\beta)$ is larger than $M_{\beta}$ for $\epsilon \Theta > 0$.
Figure~\ref{Fig:Theta_Phim:B1.0-10_L0-10} indicates that 
$M(q,\beta)$ at $q \neq 1$ is smaller than $M_{\beta}$
for small and positive $\epsilon$ in the large area of the parameters, 
where $\Theta$ is negative.
This tendency is probable intuitively.
The cutoff exists in Tsallis statistics of $q < 1$ ($\epsilon > 0$),  
then the fluctuation should be smaller than that in BG statistics.  
Therefore $\langle x^{2} \rangle$ in Tsallis statistics of $q<1$ should be smaller than 
that in BG statistics.  
Then $M(q<1,\beta)$ should be smaller than $M_{\beta}$. 
However $\Theta$ is positive for small $\beta_p$, and then 
$M(q<1,\beta)$ is larger than $M_{\beta}$ even when $\epsilon$ is positive.
In addition, the modification of $\varphi_{\rm tot}$ due to the deviation from $q=1$
is at most $O\left( \left| \epsilon \varphi_m \right| \right)$ 
in the large area of this figure, 
because the absolute value of $\Theta$ is less than that of $\varphi_m$.

\begin{figure}
\subfigure[The value of $\varphi_m$]{
\includegraphics[width=0.45\textwidth]{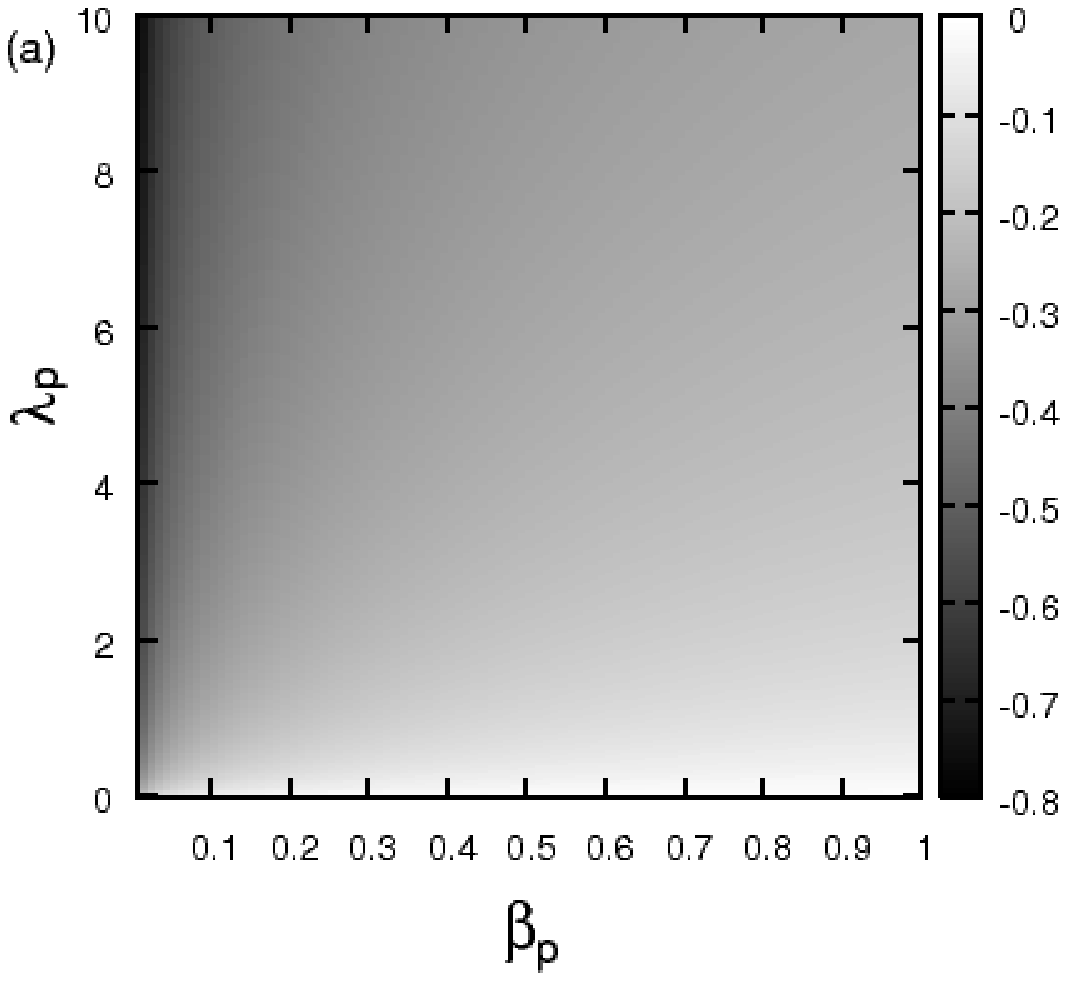}
\label{Fig:Phim_B0.01-1_L0-10_Thermal}
}
\subfigure[The value of $\Theta$]{
\includegraphics[width=0.45\textwidth]{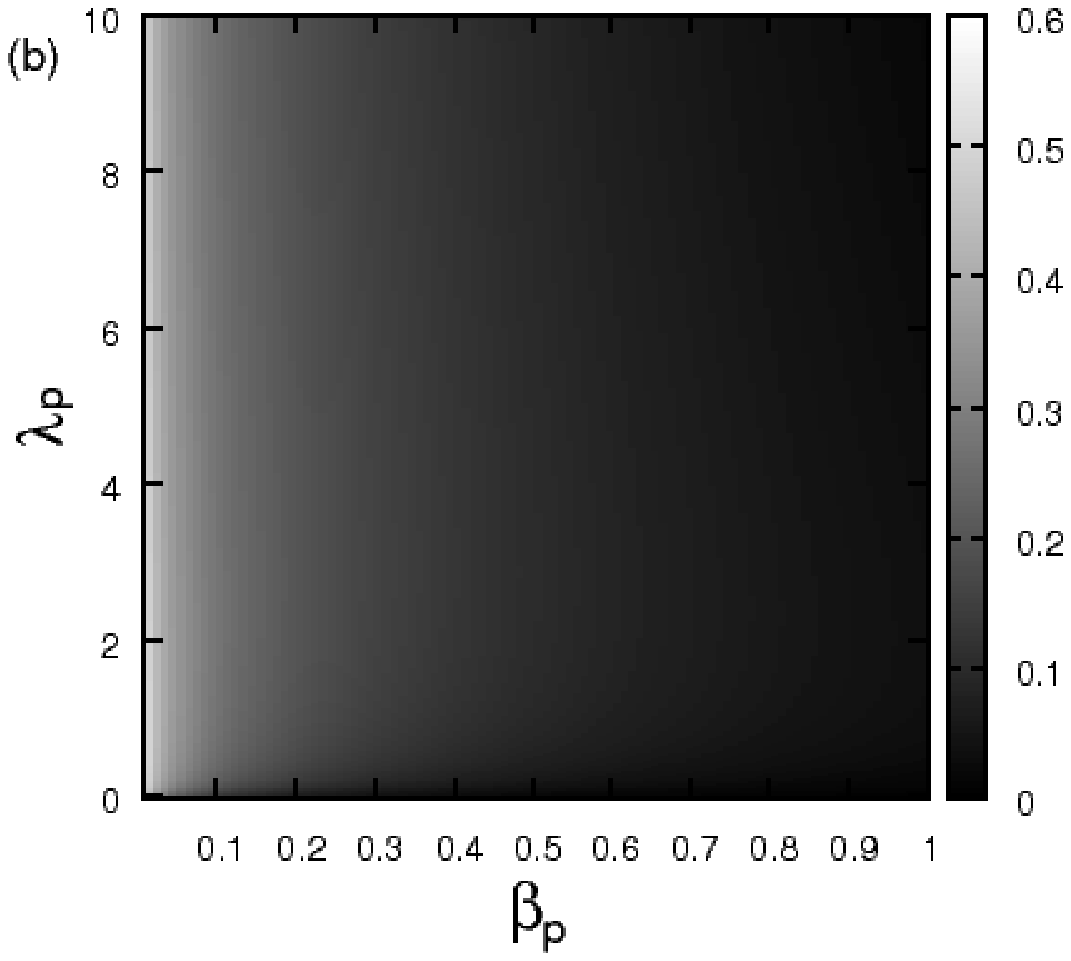}
\label{Fig:Theta_B0.01-1_L0-10_bk}
}
\caption{The squeeze angle in the ranges of $0.01 \le \beta_p \le 1$ and $0 \le \lambda_p \le 10$.}
\label{Fig:Theta_Phim:B0.01-1_L0-10}
\end{figure}

The absolute value of $\Theta$ becomes large as $\beta_{p}$ goes to zero 
in Fig.~\ref{Fig:Theta_Phim:B1.0-10_L0-10}(b).
Therefore I show the map of $\Theta$ at small $\beta_p$. 
Figure~\ref{Fig:Phim_B0.01-1_L0-10_Thermal} is the map of $\varphi_m$ in the ranges of  
$0.01 < \beta_p < 1$ and $0 < \lambda_p < 10$.
Figure~\ref{Fig:Theta_B0.01-1_L0-10_bk} is the map of $\Theta$ in the same ranges.
The absolute value of $\Theta$ is close to that of $\varphi_m$ in this region.
Therefore the difference between BG statistics and Tsallis statistics 
will be important.
The modification due to $q \neq 1$ is given as $-\epsilon \Theta$, 
and then the ratio of modification is $O\left(\left| \epsilon \right|\right)$
in the present case. 
The quantity $M(q,\beta)$ is larger than $M_{\beta}$ for $\epsilon > 0$,
because the sign of $\Theta$ is positive at small $\beta_p$.

From Fig.~\ref{Fig:Theta_Phim:B1.0-10_L0-10} and Fig.~\ref{Fig:Theta_Phim:B0.01-1_L0-10}, 
the effects of the discrepancy between $q \neq 1$ and $q=1$ is relatively large 
at small $\lambda_p$ (weak coupling) and small $\beta_p$ (high temperature).
Therefore the experiment to measure the deviation from the Boltzmann-Gibbs statistics
should be performed at weak coupling and high temperature 
in the system described by the $x^{4}$ potential.

\section{Discussion and Conclusion}
\label{sec:conclusions}
I studied the effects of the environments described by Tsallis statistics in quantum mechanics,
when the deviation from Boltzmann-Gibbs (BG) statistics is small.  
I calculated the squeeze angle (the angle in Bogoliubov transformation)
in the mean approximation. 
I used the $x^{4}$ model, and 
the effects of $q \neq 1$ in Tsallis statistics were taken into 
the coefficient of the $x^2$ term through the expectation value $\langle x^{2} \rangle_{q}$.
The squeeze angles were displayed for various sets of the dimensionless parameters,
the inverse of the temperature $\beta_p$ and the coupling strength $\lambda_{p}$.

In the numerical calculations, I showed that 
the sign of the squeeze angle $\Theta$ changes as a function of $\beta_p$ and $\lambda_p$.
Then the sign of $\theta(q,\beta) = - \epsilon \Theta(\beta)$ is not determined by 
only the sign of the parameter $\epsilon$ which is defined by $1-q$,
where this parameter indicates the property of the statistics, subextensive or superextensive.
Intuitively, 
it seems that the expectation value $\langle x^{2} \rangle_{q=1-\epsilon}$ is large 
in the superextensive case. 
However, the fact that $\Theta$ changes the sign due to the values, $\beta_p$ and $\lambda_p$, 
indicates that this conjecture is not always correct.

The squeeze angle as a function of $\beta_p$ has a dip structure, 
and this structure is easily seen at large $\lambda_p$.
This structure is characteristic, 
because the angle due to the thermal effect is a monotone function. 
Considering the ratio of the squeeze angle caused by the difference 
between BG statistics and Tsallis statistics 
to the squeeze angle caused by BG static (thermal contribution), 
I found that the absolute value of this ratio at high temperature and weak coupling
is relatively large in the ranges of the numerical calculations.
From these facts, 
the effects caused by the difference between Tsallis statistics and BG statistics 
will be observed at high temperature. 

The effect of the environment was included in the quantity $M(q,\beta)$.
The $x^2$ term in eq.~\eqref{eqn:MeanFieldHamiltonian}
can be rewritten as follows:
\begin{equation}
\frac{1}{2m} M^{2}(q,\beta)\  \omega^2 x^2 = 
\frac{1}{2m} m^2 \left( \frac{M^{2}(q,\beta)}{m^2} \  \omega^2 \right) x^2
\label{eqn:modulation}
\end{equation}
The right-hand side of eq.~\eqref{eqn:modulation} is interpreted as 
 the modulation of the frequency of a particle with mass $m$. 
This effect may be found experimentally in a quantum mechanical system.
 
I studied the effects of the environment described by Tsallis statistics in this paper. 
I hope that this work is useful in the future studies of the non-extensive statistic.

\section*{Acknowledgements}
The author thanks Dr.~Horikoshi and Dr. Osada for useful comments.

\bibliographystyle{elsarticle-num}
\bibliography{mypaper}
\end{document}